\documentclass[10pt]{article}
\usepackage[OE]{express}
\usepackage{amsmath}

\begin{document}
\title{Concurrency of anisotropy and spatial dispersion in low refractive index dielectric composites}

\author{Andrey A. Ushkov,\authormark{1} and Alexey A. Shcherbakov\authormark{1,*}}

\address{\authormark{1}Laboratory of Nanooptics and Plasmonics, Moscow Institute of Physics and Technology, 9 Institutsky, Dolgoprudnyi, Moscow Region 141700, Russia}

\email{\authormark{*}alex.shcherbakov@phystech.edu} 



\begin{abstract}
The article demonstrates uncommon manifestation of spatial dispersion in low refractive index contrast 3D periodic dielectric composites with periods of about one tenth of the wavelength. First principles simulations by the well established plane wave method reveal that spatial dispersion leads to appearance of additional optical axes and can compensate anisotropy in certain directions.
\end{abstract}

{\small One print or electronic copy may be made for personal use only. Systematic reproduction and distribution, duplication of any material in this paper for a fee or for commercial purposes, or modifications of the content of this paper are prohibited.}

\url{https://www.osapublishing.org/oe/abstract.cfm?URI=oe-25-1-243}



\section{Introduction}

Spatial dispersion effects in elastic light interaction with 3D periodic structures result in various phenomena unusual to isotropic media, and strongly influence propagation of surface waves \cite{Agranovich1966}. Currently manifestations of the spatial dispersion are well-known both for natural crystals \cite{Agranovich1966,Levine2001} and artificial periodic structures being either photonic crystals or metamaterials \cite{Rumpf2015}.

Nonlocality in natural crystals is weak at optical wavelengths, and results in directional variations of mode propagation constant at scales of about $10^{-5} - 10^{-6}$ \cite{Ginzburg1958}. However, cubic crystals appear to possess seven optical axes, when such deviations are considered. This effect called spatial-dispersion-induced birefringence, was already known by Lorentz \cite{Lorentz1936}, and nowadays is taken into account by manufacturers of projection systems for deep ultraviolet lithography set-ups \cite{Serebriakov2004,Levine2002}.

On the other hand, nonlocality in artificial photonic structures -- photonic crystals and metamaterials -- is much more prominent \cite{Rumpf2015,Vynck2009} and draws a lot of attention due to promising applications. Strong spatial dispersion in photonic crystals results in superprism effect, possibility for perfect imaging, and generally provides advanced potential for spatial filtering and beam manipulation \cite{Notomi2010,Maigyte2013}. Interpretation of phenomena related to spatial dispersion appears to be handy with representation by isofrequency contours and surfaces \cite{Kosaka2000,Notomi2002,Baba2002,Loiko2007,Maigyte2015}. For certain classes of metamaterials isofrequency contours were demonstrated to be dramatically distorted in comparison with those of natural crystals highly depending on structure parameters \cite{Gorlach2015,Fietz2011}. Spatial-dispersion-induced birefringence being small for natural materials appeared to have prominent values for metamaterials with resonant metallic inclusions \cite{Chebykin2015,Gorlach2016}. Authors of \cite{Koshelev2016} demonstrated a theoretical possibility to stop the light in metamaterial waveguide coming from an interplay between the spatial dispersion and anisotropy.

An intermediate position between weak spatial dispersion in natural crystals and prominent effects in photonic crystals and metamaterials can be filled with low refractive index periodic dielectric structures of period-to-wavelength ratio of the order of $0.1$, so that no diffraction occurs. Paper \cite{Netti2001} provided an evidence that such composites can exhibit quite unusual optical properties by demonstrating 2D periodically patterned waveguide to manifest trirefringence. Current technology level of the direct laser writing technique allows fabricating even 3D periodic high-quality dielectric composites with periods as small as tenths of a micron for operating at mid-infrared and telecom wavelengths \cite{Maruo2008}. To estimate an order of magnitude of spatial dispersion effects in such structures consider permittivity decomposition under assumption that structure is non-gyrotropic:
\begin{equation}
  \varepsilon_{\alpha\beta}({\bf{q}}) = \varepsilon_{\alpha\beta}^{(0)} + \zeta_{\alpha\beta\gamma\delta}q_{\gamma}q_{\delta} + \ldots,
  \label{eq:one}
\end{equation}
where $\bf{q}$ is wavevector. The second term scales as $(\Lambda/\lambda)^2$ with $\Lambda$ and $\lambda$ being characteristic period of a composite and the wavelength respectively \cite{Agranovich1966}. For structures of interest set $\Lambda/\lambda\sim 10^{-2}-10^{-1}$, which yields $|\varepsilon^{-1}-(\varepsilon^{-1})^{(0)}|\sim 10^{-4}-10^{-2}$. Taking into account that optical anisotropy in such structures in the limit of infinitely small period amounts to $10^{-3}-10^{-2}$ for low index dielectrics and to $10^{-1}$ for high index dielectrics \cite{Shcherbakov2015,Che2007}, it becomes clear, that spatial dispersion terms can compete with anisotropy of the first term $\varepsilon_{\alpha\beta}^{(0)}$, which may lead to interesting effects for low refractive index contrast composites operating away from resonances.

\section{Plane wave expansion method}
In order to investigate spatial dispersion in the described parameter range we utilized a variation of the plane wave expansion method for first principle simulation of photonic structures \cite{Leung1990,Ho1990,Johnson2001}. By using such first principle approach, from the one hand, we avoid restricting ourselves by decomposition of Eq.~(\ref{eq:one}) and consequent loss of accuracy due limited applicability of Eq.~(\ref{eq:one}) for finite period structures. From the other hand, this method provides means for estimation of a range of applicability of truncated series in Eq.~(\ref{eq:one}). The plane wave expansion method can be formulated on the basis of both differential and integral equations. In the current research we use the electric field formulation, which is based on the volume integral equation solution of the Maxwell's equations \cite{Morse1953}
\begin{equation}
  E_{\alpha}({\bf{r}}) = E^{inc}_{\alpha}({\bf{r}})
	+ i\omega\mu_0 \int\limits_{V'} { \left(\delta_{\alpha\beta} + \frac{1}{k_b^2}\frac{\partial}{\partial x_{\alpha}}\frac{\partial}{\partial x_{\beta}}\right)\frac{\exp\left(ik_b\left|{\bf{r}} - {\bf{r'}}\right|\right)}{4\pi\left|{\bf{r}} - {\bf{r'}}\right|} J_{\beta}({\bf{r'}}) dV' },
  \label{eq:two}
\end{equation}
where $k_b^2=\omega^2\varepsilon_b\mu_0$ ($\varepsilon_b$ is some constant, and $\mu_0$ is the permeability of vacuum), $\alpha,\beta=1,2,3$ enumerate Cartesian axes, $\delta_{\alpha\beta}$ is the Kronecker symbol, and the multiplier before source $J_{\beta}({\bf{r'}})$ is the free space dyadic Green function. $E^{inc}_{\alpha}({\bf{r}})$ denotes components of some pre-defined incident field, and $E_{\alpha}({\bf{r}})$ is unknown field. Taking the source as to be generated by spatial permittivity changes ${\bf{J}}=-i\omega(\varepsilon-\varepsilon_b){\bf{E}}$ one obtains a self-consistent equation on the unknown electric field. In the discreet 3D Fourier space Eq.~(\ref{eq:two}) becomes
\begin{equation}
  E_{\alpha{\bf{m}}} - \frac{k_b^2\delta_{\alpha\beta} - k_{\alpha{\bf{m}}}k_{\beta{\bf{m}}}}{k^2_{\bf{m}} - k^2_b} \left(\frac{\Delta\varepsilon}{\varepsilon_b}\right)_{{\bf{m}}-{\bf{m'}}} E_{\beta{\bf{m'}}} = E^{inc}_{\alpha{\bf{m}}}.
  \label{eq:three}
\end{equation}
Here vector index ${\bf{m}}=(m_1,m_2,m_3)$ enumerates Fourier harmonics, possible ${\bf{k}}$ component values are $k_{\alpha{\bf{m}}}=k_{\alpha}^{inc}+m_{\alpha}K_{\alpha}$, $K_{\alpha}=2\pi/\Lambda_{\alpha}$, and $\Lambda_{\alpha}$ are periods of the composite structure along coordinate axes. Notation for wavevectors here is different from Eq.~(\ref{eq:one}) to distinguish wavevectors $\bf{q}$ of propagating modes. Propagation constants and polarizations of eigen modes in a given 3D periodic composite can be retrieved from solution of either an eigenvalue problem coming from Eq.~(\ref{eq:three}) with zero incident field $E^{inc}_{\alpha{\bf{m}}}$, or a resonant analysis of Eq.~(\ref{eq:three}). We implemented the latter possibility through the use of an efficient complex pole search algorithm \cite{Shcherbakov2015}.

Special precautions were taken to control simulation accuracy at each step of the algorithm, as we are interested in effects which can occur for relatively small effective refractive index contrasts starting from values of about $10^{-4}$. To satisfy this requirement we first set a sufficiently low convergence criterion for the Generalized Minimal Residual method for solution of the linear equation system (\ref{eq:three}), and, second, apply the second order Richardson extrapolation to improve calculated propagation constants and harmonic field amplitudes for subsequent numbers of Fourier harmonics used in simulations. This allowed obtaining mode propagation constants with 5-7 significant digit precision.

\section{Spatial dispersion effects}

\begin{figure}
\centering\includegraphics[width=11cm]{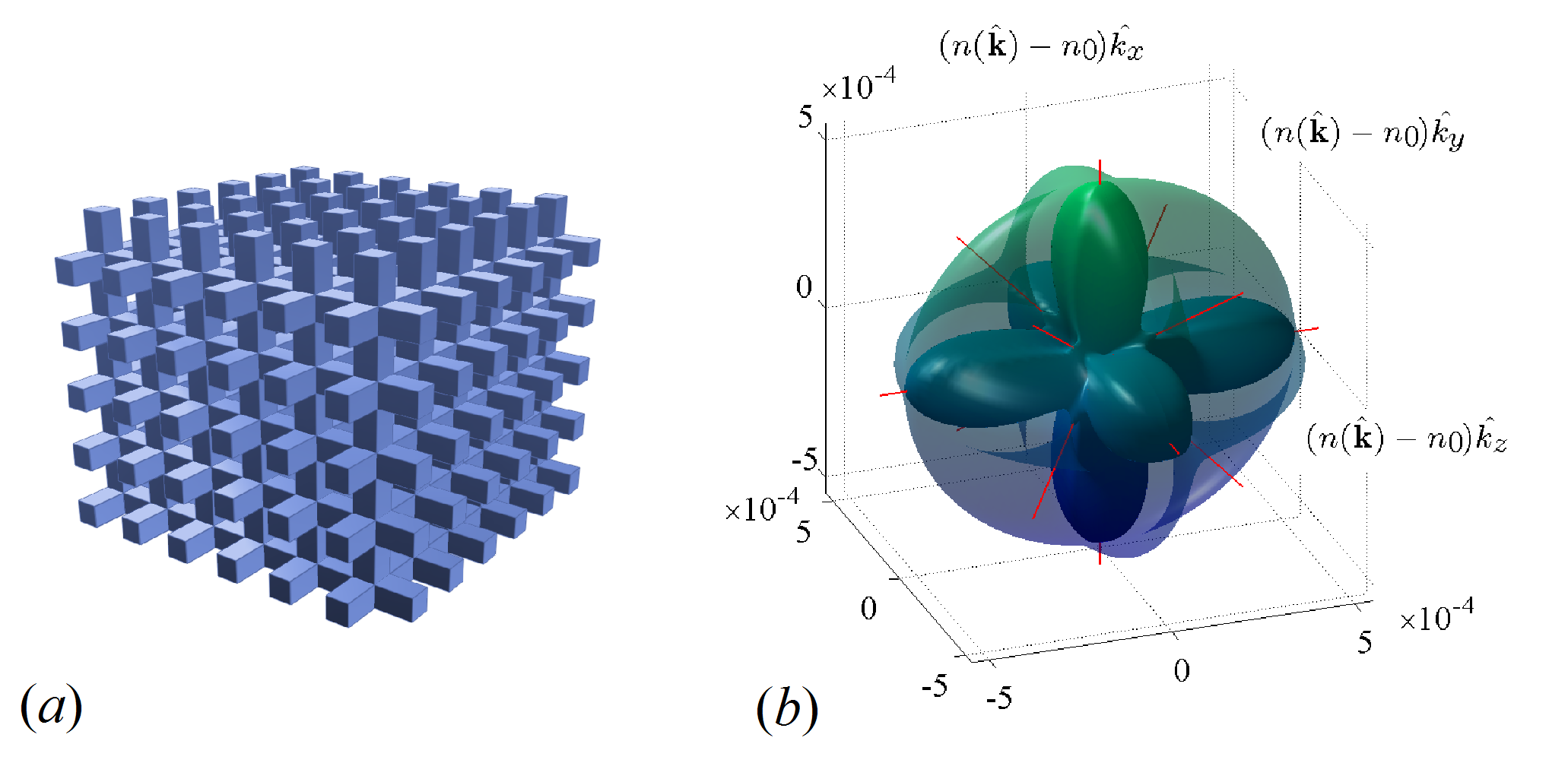}
\caption{Considered scaffold structure (a); and isofrequency surface (b) for $\Lambda/\lambda = 0.1$ possessing seven optical axes marked with red lines.}
\label{fig:1}
\end{figure}

In what follows we provide three illustrative examples of spatial dispersion manifestation. First example concerns the above mentioned existence of seven optical axes in cubic crystals, and aims at quantitatively estimating an impact of the second and higher order terms in Eq.~(\ref{eq:one}). Consider a scaffold structure shown in Fig.~\ref{fig:1}(a). For all examples let the refractive index of constituting rods be $n_c = 1.6$, rod size-to-period ratio be $1/3$, and surrounding medium be the air. Figure~\ref{fig:1}(b) shows isofrequency surfaces for two propagating modes and directions of optical axes in crystal with cubic lattice and $\Lambda/\lambda=0.1$. In order to clearly represent small differences we choose the scale along each axis $X_{\alpha}$ to be $[n(\hat{\bf{s}})-n_0]\hat{s}_{\alpha}$, with $n_0$ being the same constant for all three Cartesian axes, and $n(\hat{\bf{s}})$ being calculated propagation constant in direction $\hat{\bf{s}} = {\bf{q}}/|{\bf{q}}|$.

\begin{figure}
\centering\includegraphics[width=12cm]{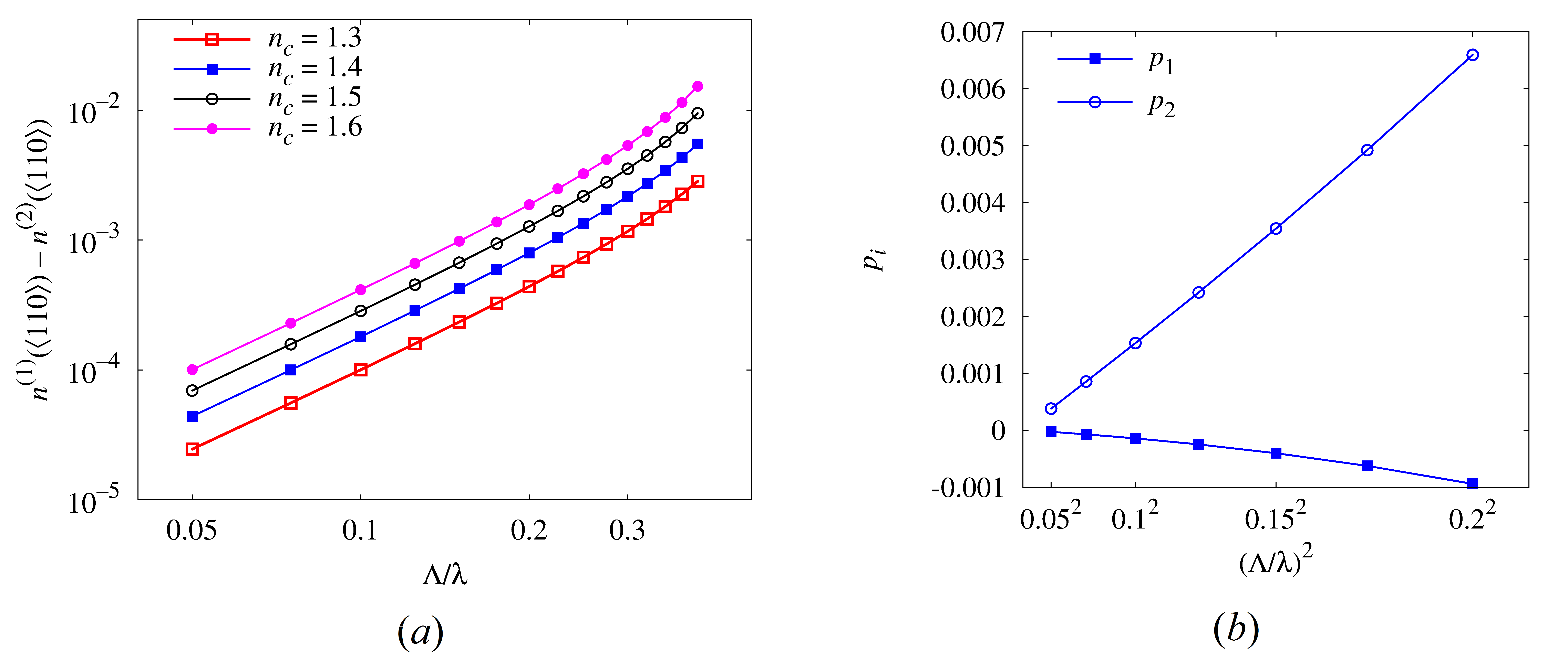}
\caption{(a) Dependence of maximum effective refractive index difference of two modes propagating in $\langle110\rangle$ direction from $\Lambda/\lambda$ ratio in doubly logarithmic scale. Four curves correspond to different refractive indices of a dielectric material constituting the scaffold structure shown in Fig.~\ref{fig:1}(a) with $\Lambda_1 = \Lambda_2 = \Lambda_3$. (b) Dependence of coefficients $p_{1,2}$ from $(\Lambda/\lambda)^2$ in Eq.~(\ref{eq:four}).}
\label{fig:2}
\end{figure}

Dependence of absolute values of maximum effective refractive index difference for two propagating modes in the structure, which occurs in $\langle110\rangle$ directions, from the period-to-wavelength ratio is shown in Fig.~\ref{fig:2}(a) in doubly logarithmic scale. The curves reveal that series truncation by second order terms in Eq.~(\ref{eq:one}) is valid roughly up to periods about $\Lambda/\lambda\approx 0.2$ independently of composite dielectric constant within low contrast range $n_c-1\sim0.5$. Then, for periods lower than $0.2$ we can extract numerical values of coefficients before second order terms in permittivity decomposition. For cubic crystals there are only three independent components of tensor $\zeta_{\alpha\beta\gamma\delta}$, namely, $\zeta_{1111}$, $\zeta_{1212}$, and $\zeta_{1122}$ \cite{Nye1985}. This permits reducing Eq.~(\ref{eq:one}) to \cite{Levine2002,Nye1985}:
\begin{equation}
  \varepsilon_{\alpha\beta} = \varepsilon^{(0)}\delta_{\alpha\beta} + n^2(\hat{\bf{s}}) \left\{ \left[p_2 + (p_1-p_2)\hat{s}_{\alpha}^2\right]\delta_{\alpha\beta} + p_3\hat{s}_{\alpha}\hat{s}_{\beta} \right\}
  \label{eq:four}
\end{equation}
with parameters $p_1$, $p_2$, $p_3$ depending on $\Lambda/\lambda$. Our simulations do not allow to account for longitudinal term $n^2p_3\hat{s}_{\alpha}\hat{s}_{\beta}$ since its multiplication by the electric field yields values of order of $10^{-5}$ at maximum (because $p_3\sim(\Lambda/\lambda)^2\sim10^{-2}-10^{-3}$, and $|\pi/2-\angle(\hat{\bf{s}},{\bf{E}})|<10^{-3}$). So we put $p_3 = 0$. The other two coefficients $p_{1,2}$ are obtained from the Helmholtz equation $n^2(\hat{\bf{s}})\,\hat{\bf{s}}\times\hat{\bf{s}}\times{\bf{E}} + \varepsilon{\bf{E}} = 0$ by substitution of calculated mode propagation constants $n(\hat{\bf{s}})$. This yields the expected quadratic dependence of $p_{1,2}$, as Fig.~\ref{fig:2}(b) shows, with $p_1 = -0.014(\Lambda/\lambda)^2$, and $p_2 = 0.153(\Lambda/\lambda)^2$. Value $\varepsilon^{(0)} = 1.31866$ was precalculated as a limit of effective refractive index along coordinate axis at $\Lambda/\lambda = 0$ under assumption of its quadratic behaviour. Besides, one can calculate decomposition of inverse permittivity $\varepsilon^{-1}$ similarly. This example also demonstrates that maximum propagation index difference in cubic crystals practically does not exceed $10^{-2}$ when the spatial dispersion is described by the second order term for structures of interest.

\begin{figure}
\centering\includegraphics[width=11cm]{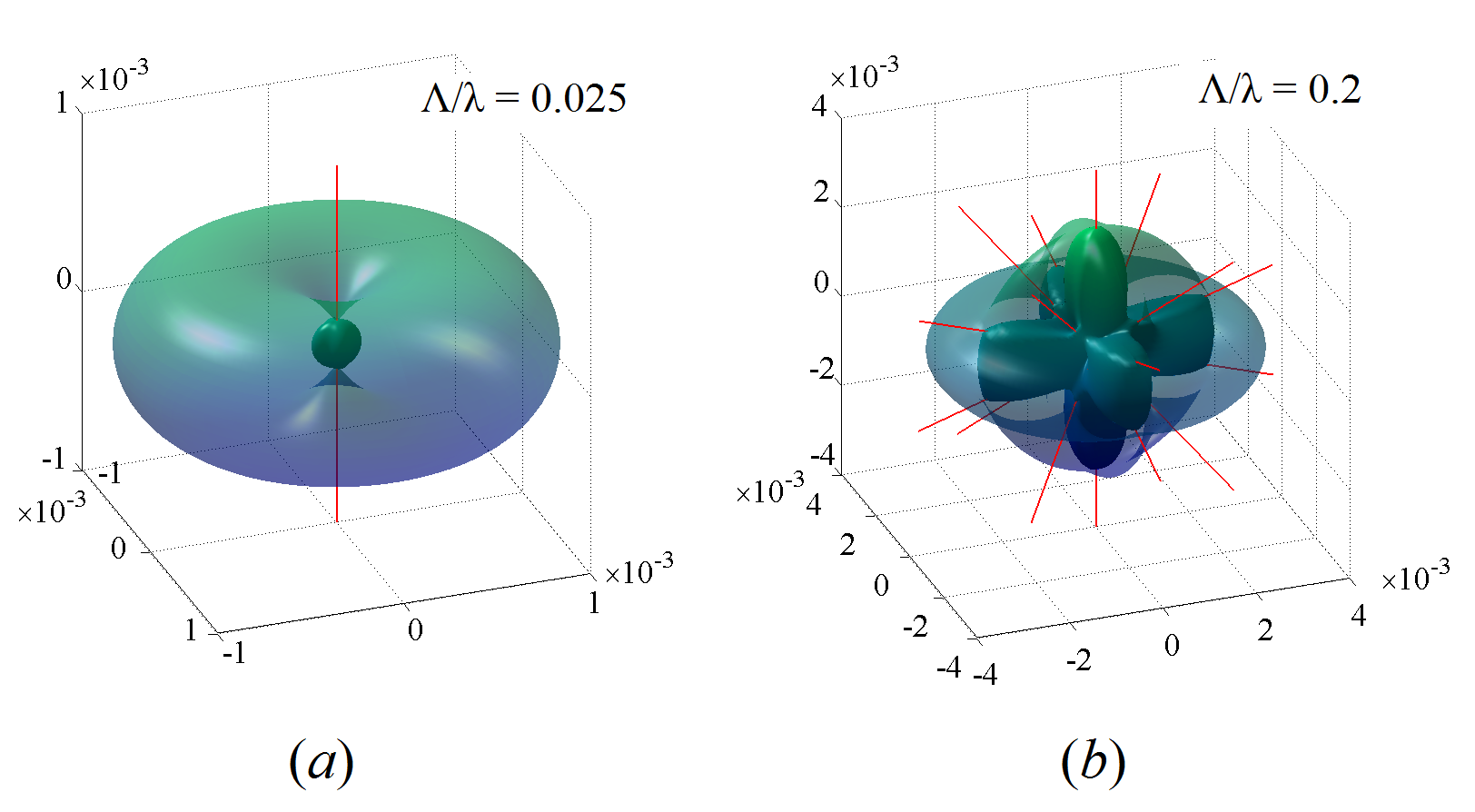}
\caption{(a) Isofrequency surface of effectively uniaxial crystal for small period value $\Lambda_1/\lambda = 0.025$. (b) Distortion of the uniaxial crystal isofrequency surface and position of 8 additional optical axes for period to wavelength ratio $\Lambda_1/\lambda = 0.2$ (see {\bf Visualization 1} for sequential period change). Considered scaffold structure shown in Fig.~\ref{fig:1}(a) has tetragonal lattice with period relations $\Lambda_2 = \Lambda_1$, $\Lambda_3 = 1.1\Lambda_1$. Axis scales are the same as in Fig.~\ref{fig:1}(b).}
\label{fig:3}
\end{figure}

For the second example modify the considered scaffold structure so as one of periods to be different from the two others: $\Lambda_1 = \Lambda_2$, $\Lambda_3 = 1.1\Lambda_1$ and consider tetragonal lattice. In this case we get effectively uniaxial medium in the limit $\Lambda_1\rightarrow 0$ with optical axis coinciding with the third Cartesian axis. Figure~\ref{fig:3}(a) demonstrates the well-known refractive index sphere and ellipsoid in the above described coordinates. Increase of periods results in appearance of four additional optical axes in planes $(110)$ analogously to the first case of cubic lattice, and another four optical axes in planes $(100)$ and $(010)$, which come from splitting of cubic lattice axes in directions $\langle100\rangle$ and $\langle010\rangle$ due to symmetry reduction. These additional axes rise from $\langle001\rangle$ direction at infinitely small periods. Then, as $\Lambda_1/\lambda$ increases, axes of the first set lying in planes $(110)$ tend to approach $\langle 111\rangle$ directions, while axes of the second set approach $\langle 100\rangle$ and $\langle 010\rangle$. This means that spatial dispersion terms tend to compensate anisotropy in $\langle100\rangle$ and $\langle010\rangle$ directions.

\begin{figure}
\centering\includegraphics[width=12cm]{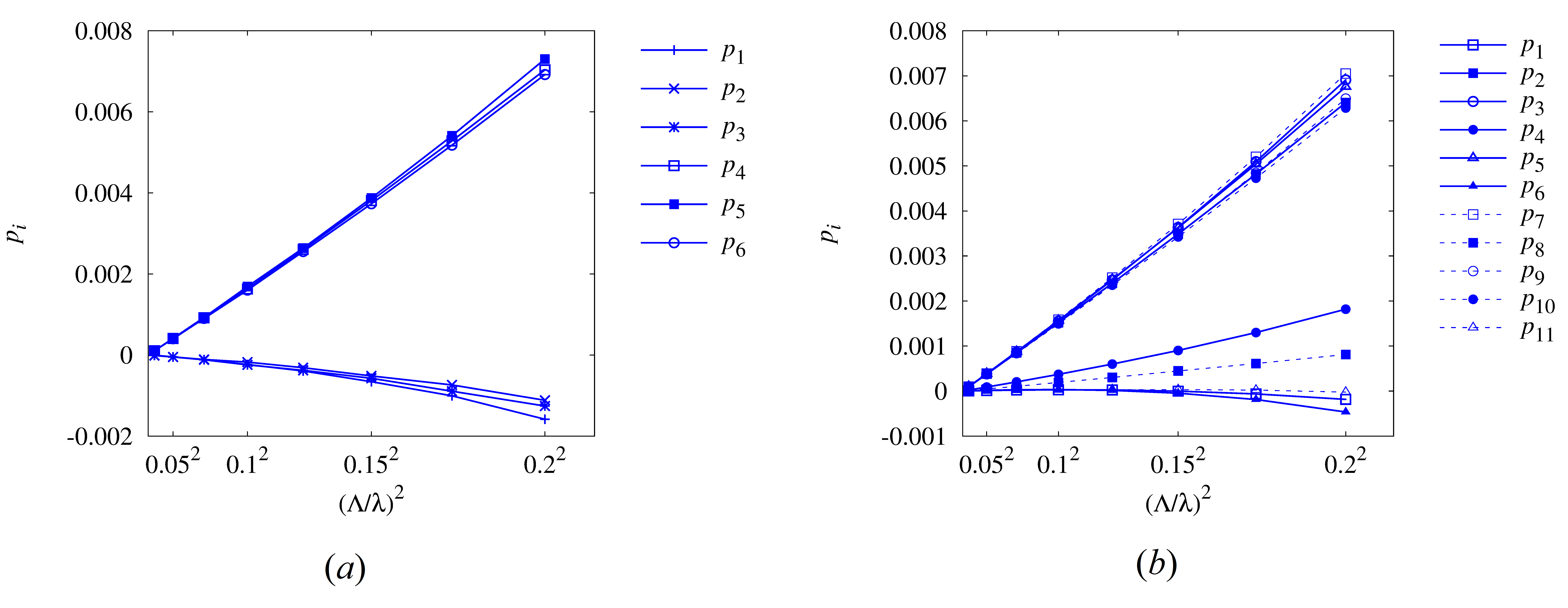}
\caption{Dependence of parameters, which define spatial dispersion terms in Eqs.~(\ref{eq:five}, \ref{eq:six}), from $(\Lambda/\lambda)^2$ for (a) tetragonal lattice and (b) orthorhombic lattice.}
\label{fig:4}
\end{figure}

In case of tetragonal lattice tensor $\zeta_{\alpha\beta\gamma\delta}$ has seven independent components \cite{Nye1985}, and one of them defines the longitudinal input. Therefore, six constants $p_1,\ldots,p_6$ define the permittivity tensor without the longitudinal term:
\begin{equation}
\begin{split}
  \varepsilon &=
  \begin{pmatrix}
  \varepsilon^{(0)}_{11}&0&0\\ 0&\varepsilon^{(0)}_{11}&0 \\ 0&0&\varepsilon^{(0)}_{33}  
  \end{pmatrix} \\
  &+ n^2(\hat{\bf{s}})
  \begin{pmatrix}
  p_1s_1^2+p_4s_2^2+p_5s_3^2&p_2s_1s_2&0 \\
  p_2s_1s_2&p_4s_1^2+p_1s_2^2+p_5s_3^2&0 \\
  0&0&p_6(s_1^2+s_2^2)+p_3s_3^2  
  \end{pmatrix}
\end{split}
  \label{eq:five}
\end{equation}
These constants are found from the Helmholtz equation similarly to the previous example. Figure~\ref{fig:4}(a) demonstrates their dependence from $(\Lambda_1/\lambda)^2$, and shows, that for periods up to $\Lambda/\lambda\sim0.15$ they write $p_1=-0.023(\Lambda/\lambda)^2$, $p_2=-0.017(\Lambda/\lambda)^2$, $p_3=-0.024(\Lambda/\lambda)^2$, $p_4=0.163(\Lambda/\lambda)^2$, $p_5=0.17(\Lambda/\lambda)^2$, and $p_6=0.16(\Lambda/\lambda)^2$. Diagonal components are $\varepsilon^{(0)}_{11} = 1.3179$, and $\varepsilon^{(0)}_{33} = 1.32019$.

\begin{figure}
\centering\includegraphics[width=11cm]{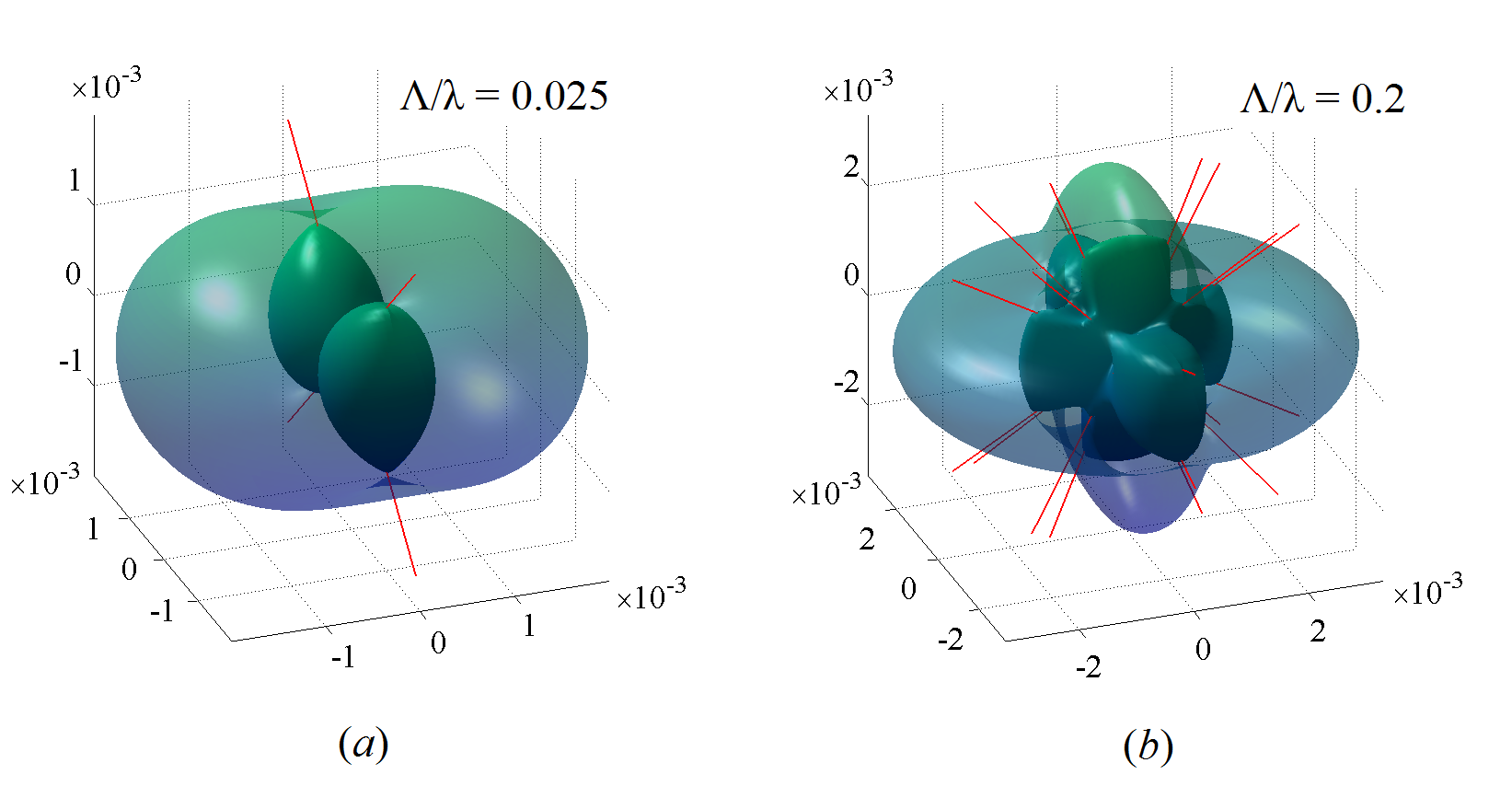}
\caption{(a) Isofrequency surface of effectively biaxial crystal for small period value $\Lambda_1/\lambda = 0.025$. (b) Distortion of the biaxial crystal isofrequency surface and position of 8 additional optical axes for period to wavelength ratio $\Lambda_1/\lambda = 0.2$ (see {\bf Visualization 2} for sequential period change). Considered scaffold structure shown in Fig.~\ref{fig:1}(a) has orthorhombic lattice with period relations $\Lambda_2 = 0.9\Lambda_1$, $\Lambda_3 = 1.1\Lambda_1$. Axis scales are the same as in Fig.~\ref{fig:1}(b).}
\label{fig:5}
\end{figure}

In the third example we put all three periods to be different $\Lambda_2 = 0.9\Lambda_1$, $\Lambda_3 = 1.1\Lambda_1$ to consider an orthorhombic lattice, so the crystal behaves as effectively biaxial medium in the limit $\Lambda_1\rightarrow 0$, which is shown in Fig.~\ref{fig:5}(a). With increase of periods there remain two optical axes up to $\Lambda_1/\lambda\approx0.15$. At larger values of $\Lambda_1/\lambda$ the lattice isofrequency surface transits to a state with 10 optical axes. In comparison with the first example the rotational symmetry is broken along all coordinate axes, and consequently all three optical axes of the cubic crystal are split in two at sufficiently large periods (Fig.~\ref{fig:5}(b)). Longitudinal term being extracted, permittivity tensor tensor writes in this case via 11 parameters:
\begin{equation}
\begin{split}
  \varepsilon &=
  \begin{pmatrix}
  \varepsilon^{(0)}_{11}&0&0\\ 0&\varepsilon^{(0)}_{22}&0 \\ 0&0&\varepsilon^{(0)}_{33}  
  \end{pmatrix} \\
  &+ n^2(\hat{\bf{s}})
  \begin{pmatrix}
  p_1s_1^2+p_2s_2^2+p_3s_3^2 & 0 & p_4s_1s_3 \\
  0 & p_5s_1^2+p_6s_2^2+p_7s_3^2 & p_8s_2s_3 \\
  p_4s_1s_3 & p_8s_2s_3 & p_9s_1^2+p_{10}s_2^2+p_{11}s_3^2  
  \end{pmatrix}
\end{split}
  \label{eq:six}
\end{equation}
Dependence of these parameters from $\Lambda_1/\lambda$ ratio is shown in Fig.~\ref{fig:4}(b). Corresponding numerical coefficients for small period values $\Lambda_1/\lambda<0.15$ write $p_1=-0.003(\Lambda/\lambda)^2$, $p_2=0.152(\Lambda/\lambda)^2$, $p_3=0.155(\Lambda/\lambda)^2$, $p_4=0.037(\Lambda/\lambda)^2$, $p_5=0.157(\Lambda/\lambda)^2$, $p_6=-0.003(\Lambda/\lambda)^2$, $p_7=0.158(\Lambda/\lambda)^2$, $p_8=0.019(\Lambda/\lambda)^2$, $p_9=0.151(\Lambda/\lambda)^2$, $p_{10}=0.15(\Lambda/\lambda)^2$, and $p_{11}=-0.003(\Lambda/\lambda)^2$ with diagonal components $\varepsilon^{(0)}_{11} = 1.31887$, $\varepsilon^{(0)}_{22} = 1.316136$, and $\varepsilon^{(0)}_{33} = 1.321055$.

\section{Conclusion}
To conclude, our work reveals existence of nontrivial spatial dispersion effects in low refractive index contrast 3D periodic dielectric composites operating in the effective medium regime. For sufficiently large period values artificial crystals possessing uniaxial and biaxial properties in the limit of infinitely small periods are demonstrated to have $9$ and $10$ optical axes respectively. In addition, quantitative values of effective parameters of composites retrieved from results of first-principle simulations yield a tendency for the spatial dispersion to compensate anisotropy in certain directions. These effects can be regarded as strong in a sense that spatial dispersion terms have the same order of magnitude as anisotropy in the considered structures.

We verified that small absorption ($\operatorname{Im}(n)\lesssim 10^{-2}$) in a dielectric material constituting a composite does not qualitatively affect the results, and only leads to appearance of proportionally small imaginary parts of effective refractive indices. The described effects may be used in applications which require advanced spectral and spatial filtering of the electromagnetic radiation. One may expect, that a plane wave interacting with the investigated structures should be tolerant to small defects inevitably present in experimental samples: due to small period-to-wavelength ratio the wave phase appreciably changes at scales of dozens of periods, thus, making only average period filling and length to play a role. 

Presented retrieval of $\zeta_{\alpha\beta\gamma\delta}$ values is a kind of a numerical homogenization procedure for periodic dielectric composites with determined range of validity. This procedure is a direct use of effective propagation constants and corresponding polarizations, obtained through the first-principles calculations, in the Maxwell's equations under assumption of validity of decomposition in Eq.~(\ref{eq:one}). Probably, analogous results can be obtained from various procedures elaborated within the homogenization theory, which is still being intensively developed (e.g., see review \cite{Simovski2011} and references therein). However, we did not intend to evaluate all possible approaches. Main purpose of the paper is to demonstrate the effects appearing in periodic structures and to quantitatively characterize them.

\section*{Funding}
Russian Ministry of Education and Science (16-19-2014/K); Russian Foundation for Basic Research (16-29-11747-ofi-m and 16-07-00837).

\end{document}